\documentclass[ACS]{WileyNJD-v1}
\usepackage{amsfonts}
\usepackage{amsmath}
\usepackage{amssymb}
\usepackage{graphicx}%
\providecommand{\U}[1]{\protect\rule{.1in}{.1in}}

\newcommand\symtextcomp{0}

\articletype{Original Article}%

\begin{document}

\title{On the Kubo-Greenwood Model for Electron Conductivity}

\author[1]{James Dufty}

\author[1]{Jeffrey Wrighton}

\author[2]{Kai Luo}

\author[1,2]{S.B. Trickey}

\address[1]{\orgdiv{Department of Physics}, \orgname{University of Florida}, \orgaddress{\state{Florida}, \country{USA}}}

\address[2]{\orgdiv{Quantum Theory Project}, \orgname{University of Florida}, \orgaddress{\state{Florida}, \country{USA}}}

\corres{James Dufty, Department of Physics, University of Florida, Gainesville FL  USA \email{dufty@phys.ufl.edu}}

\presentaddress{Department of Physics, University of Florida, P.O. Box 118435, Gainesville FL 32611-8435, USA}

\abstract[Abstract]{Currently, the most common method to calculate transport properties for
materials under extreme conditions is based on the phenomenological Kubo-Greenwood method. The results of an inquiry into the justification and context
of that model are summarized here. Specifically, the basis for its connection
to equilibrium DFT and the assumption of static ions are discussed briefly.}



\maketitle

\section{Introduction and Motivation} \label{sec1}

Recent interest in matter under extreme conditions provides
motivation for new theoretical methods to describe the thermodynamic and
transport properties of such systems. The relevant conditions include those of high
pressure materials, warm dense matter, and high temperature plasmas
\cite{Glenzer16}. The generic example is a system of electrons and ions at
equilibrium. Thermodynamic properties such as pressure and free energy are
typically described via \textit{ab initio} molecular dynamics (AIMD) simulation
 for the classical ions, with the effective forces due to the electrons
calculated from finite temperature density functional theory (DFT)
\cite{AIMD}.
Significant progress has been made within DFT for this
purpose. In contrast, the exploitation of equilibrium DFT results,
in particular Kohn-Sham quantities, for the
calculation of electronic transport properties is far from obvious.
Nevertheless, plausible but uncontrolled approximations are used in practice
to calculate transport properties by promoting the Kohn - Sham Hamiltonian
from a tool for
constructing the equilibrium density, to a generator for the electron
dynamics. This bold leap of faith accomplishes the objective of a calculation
incorporating the equilibrium tools of AIMD and DFT to the calculation of
transport properties \cite{KG}. The approach is known as the Kubo-Greenwood
model. The objective here is to report the results of an inquiry into the
justification and context of this approach. Two issues are addressed: the
origin of the Kohn-Sham dynamics, and the assumption of a frozen ion
configuration. Details of these results and their derivation will be presented
elsewhere. Although the discussion is focused on electron conductivity the
analysis applies also to other transport coefficients, e.g. thermal
conductivity and viscosity.

\section{Green-Kubo Conductivity and Kubo-Greenwood approximation}\label{sec2}

The starting point for the analysis is the formally exact Green -
Kubo expression for the frequency dependent electron conductivity, in terms of
the current autocorrelation function
\begin{equation}
\operatorname{Re}\sigma\left(  \omega\right)  =\frac{1}{\omega}\left(
1-e^{-\beta\omega}\right)  \operatorname{Re}\int_{0}^{\infty}dte^{i\omega
t}\psi\left(  t\right)  ,\hspace{0.25in}\psi\left(  t\right)  =\frac{1}%
{3V}\left\langle \widehat{\mathbf{J}}(t)\cdot\widehat{\mathbf{J}}\right\rangle
\label{2.1}%
\end{equation}
where $\widehat{\mathbf{J}}$ is the electron current operator, $V$ is the
volume, $\beta$ is the inverse temperature, and the brackets denote an equilibrium grand canonical average.
A two-component system
comprised of electrons and ions at equilibrium is assumed.
Units such that $\hbar=1$ are
used. The ions are taken to be classical, while the electrons are fully quantum
mechanical. Equilibrium averages are taken jointly over the states of the
electrons and ions. In the following, the electron average is performed first,
followed by the ion average. Thus
\begin{equation}
\psi\left(  t\right)  =\left\langle \psi_{e}\left(  t\right)  \right\rangle
_{i}=\sum_{N_{i}}\frac{1}{h^{3N_{i}}N_{i}!}\int\left\{
d\mathbf{R}\right\}  \left\{  d\mathbf{P}\right\}  \rho_{i}\left(  \left\{
\mathbf{R,P}\right\}  \right)  \psi_{e}\left(  t\right)  
\end{equation}
where $\left\{  \mathbf{R,P}\right\}  $ denotes the $N_{i}$ position and
momentum vectors for the ions. Also, $\rho_{i}\left(  \left\{
\mathbf{R,P}\right\}  \right)$ is the equilibrium ion density matrix defined in terms of the ion Hamiltonian including the adiabatic (Born-Oppenheimer) potential energy surface due to the electrons.   As the classical ion average can be implemented
by AIMD (ergodic hypothesis), the difficult many-body problem therefore occurs
for the electron average $\psi_{e}\left(  t\right)  $%
\begin{equation}
\psi_{e}\left(  t\right)  =\psi_{e}\left(  t\mid\left\{  \mathbf{R}\left(
t\right)  \right\}  \right)  =\frac{1}{3V}\sum_{N_{e}}Tr^{(N_{e})}%
\widehat{\rho}_{e}\left(  \left\{  \mathbf{R}\left(  t\right)  \right\}
\right)  \widetilde{\mathbf{J}}(t\mid\left\{  \mathbf{R}\left(  t\right)
\right\}  )\cdot\widehat{\mathbf{J}} .\label{2.3}%
\end{equation}
The notation $\psi_{e}\left(  t\mid\left\{  \mathbf{R}\left(  t\right)
\right\}  \right)  $ and $\widetilde{\mathbf{J}}(t\mid\left\{  \mathbf{R}%
\left(  t\right)  \right\}  )$ denotes a \textit{functional} dependence on the
history of the ion configuration $\left\{  \mathbf{R}\left(  \tau\right)
\right\}  $ for all times $\tau\leq t$. That history is generated separately
by means of an AIMD simulation. In contrast the density operator for $N_{e}$ electrons,
$\widehat{\rho}_{e}\left(  \left\{  \mathbf{R}\left(  t\right)  \right\}
\right)  $, is a function of the instantaneous ion configuration at time $t$
\begin{equation}
\widehat{\rho}_{e}\left(  \left\{  \mathbf{R}\left(  t\right)  \right\}
\right)  =e^{\beta\Omega_{e}\left(  \left\{  \mathbf{R}\left(  t\right)\right\}  \right)
}e^{-\beta\left(  H_{e}+U_{ei}\left(  \left\{  \mathbf{R}\left(  t\right)
\right\}  \right)  -\mu_{e}N_{e}\right)  }. \label{2.4}%
\end{equation}
The electron Hamiltonian $H_{e}+U_{ei}\left(  \left\{  \mathbf{R}\left(
t\right)  \right\}  \right)  $ is that for all electrons with their Coulomb
interactions for each pair and all electron - ion Coulomb interactions. The
corresponding Heisenberg time dependence for the electron current is given by
\begin{equation}
\partial_{t}\widetilde{\mathbf{J}}(t\mid\left\{  \mathbf{R}\left(  t\right)
\right\}  )=i\left[  \left(  H_{e}+U_{ei}\left(  \left\{  \mathbf{R}\left(
t\right)  \right\}  \right)  \right)  ,\widetilde{\mathbf{J}}(t\mid\left\{
\mathbf{R}\left(  t\right)  \right\}  )\right]  . \label{2.5}%
\end{equation}
The evaluation of $\psi_{e}\left(  t\mid\left\{  \mathbf{R}\left(  t\right)
\right\}  \right)  $ entails confronting the full quantum many-body problem
for the electrons in a given moving ion configuration. For the complex states
of interest here no adequate first principles theory is currently available.

A practical, plausible but phenomenological, mitigation of this formidable Green-Kubo
expression results from replacing the electron many-body Hamiltonian
by a mean field Hamiltonian comprised of a sum of independent Kohn - Sham
single particle Hamiltonians,
\begin{equation}
H_{e}+U_{ei}\left(  \left\{  \mathbf{R}\left(  t\right)  \right\}  \right)
\rightarrow\sum_{j=1}^{N_{e}}h_{KS}(j,\left\{  \mathbf{R}\right\}  ),
\label{2.6}%
\end{equation}%
\begin{equation}
h_{KS}(j,\left\{  \mathbf{R}\right\}  )=\frac{p_{j}^{2}}{2m_{e}}%
+v_{KS}(\mathbf{r}_{j},\left\{  \mathbf{R}\right\}  \mid n_{e}). \label{2.7}%
\end{equation}
Here $\mathbf{R=R}\left(  t=0\right)  $. Thus, in addition to the independent
particle approximation there is the assumption that the ions are frozen for
the duration of the electron dynamics. Each particle has an effective
interaction with all ions via the Kohn-Sham potential as a functional of the
average electron density $n_{e}\left(  \mathbf{r},\left\{  \mathbf{R}\right\}
\right)$
\begin{equation}
v_{KS}(\mathbf{r}_{j},\left\{  \mathbf{R}\right\}  \mid n_{e})=-\sum
_{i=1}^{N_{i}}\frac{Ze^{2}}{\left\vert \mathbf{r}_{j}-\mathbf{R}%
_{i}\right\vert }+\int d\mathbf{r}\frac{e^{2}}{\left\vert \mathbf{r}%
_{j}-\mathbf{r}\right\vert }n_{e}\left(  \mathbf{r},\left\{  \mathbf{R}%
\right\}  \right)  +\frac{\delta F_{xc}\left[  n_{e} \right]  }{\delta n_{e}\left(  \mathbf{r}_{j},\left\{  \mathbf{R}\right\}\right)}. \label{2.8}%
\end{equation}
The first term is the Coulomb interaction, the second is the Hartree
interaction, while the third is an effect due to the average exchange and
correlation among the electrons via the corresponding free energy
contribution, $F_{xc}\left[  n_{e}  \right]  $. Calculation of the Kohn-Sham potential and the average
electron density are the central problem of DFT, requiring determination of
the eigenvalues and eigenfunctions of $h_{KS}(i,\left\{  \mathbf{R}\right\}
)$ as well. As noted above, this is a well-developed formalism. With the
assumed independent particle Hamiltonian it is now straightforward to
calculate $\psi_{e}\left(  t,\left\{  \mathbf{R}\right\}  \right)  $ in terms
of these eigenvalues and eigenfunctions. The final average over the frozen ion
configurations then is implemented by the ergodic hypothesis and AIMD%
\begin{equation}
\psi\left(  t\right)  =\lim_{T\to\infty}\frac{1}{T}\int_{0}^{T}d\tau\psi_{e}\left(
t,\left\{  \mathbf{R}\left(  \tau\right)  \right\}  \right)  . \label{2.9}%
\end{equation}
In practice, the time average is approximated as an algebraic average of a
small number of "snapshots" of different ion configurations.

Equations (\ref{2.6}) - \ (\ref{2.9}) define the Kubo-Greenwood
approximation for (\ref{2.3}) \cite{KG}.

\section{Origin and Context of Kubo-Greenwood approximation}\label{sec3}

In the Kubo-Greenwood model the complex Coulomb interactions
among the electrons and with the ions is replaced  by independent particles.
Furthermore, their interactions with the ions occurs via the Kohn - Sham
potential which originates in the Euler equation of DFT for the equilibrium
free energy. What is the justification for, and interpretation of, this
unusual approximation? To answer this, return to the original Green-Kubo
form and write the time correlation function in a representation appropriate
for a kinetic theory evaluation%
\begin{equation}
\psi_{e}\left(  t\mathbf{\mid}\left\{  \mathbf{R}\right\}  \right)  =\frac
{1}{3V}Tr_{1}n_{e}(\mathbf{r}_{1},\left\{  \mathbf{R}\right\}  )\phi\left(
v_{1}\right)  e\mathbf{v}_{1}\mathbf{\cdot}\mathcal{J}(x_{1},t\mid\left\{
\mathbf{R}\left(  t\right)  \right\}  ),\label{3.1}%
\end{equation}%
\begin{equation}
\mathcal{J}(x_{1},t\mid\left\{  \mathbf{R}\left(  t\right)  \right\}
)=\frac{1}{n_{e}(\mathbf{r}_{1},\left\{  \mathbf{R}\right\}  )\phi\left(
v_{1}\right)  }\sum_{N_{e}}\frac{1}{\left(  N_{e}-1\right)
!}Tr_{2,..,N_{e}}\rho_{e}\left(  x_{1},..,x_{N_{e}};\left\{  \mathbf{R}%
\right\}  \right)  \mathbf{J}\left(  -t\right)  .\label{3.2}%
\end{equation}
Here $x_{i}=\left\{  \mathbf{r}_{i},\mathbf{v}_{i}\right\}  $ denotes the
position and velocity vectors for particle $i$. Since the current $\mathbf{J}$
is the sum of single particle operators $e\mathbf{v}_{i}$ it is possible to
carry out the trace formally over all particles except one, leaving the single-particle representation (\ref{3.1}). Of course, the difficult many-body
problem remains in the evaluation of $\mathcal{J}(x_{1},t\mid\left\{
\mathbf{R}\left(  t\right)  \right\}  )$. It satisfies an exact "kinetic
equation" \cite{Boercker81}%
\begin{equation}
\partial_{t}\mathcal{J}(x_{1},t\mid\left\{  \mathbf{R}\left(  t\right)
\right\}  )+Tr_{2}\mathcal{L}\left(  x_{1},x_{2},t\mid\left\{  \mathbf{R}%
\left(  t\right)  \right\}  \right)  \mathcal{J}(x_{2},t\mid\left\{
\mathbf{R}\left(  t\right)  \right\}  )=0,\label{3.3}%
\end{equation}
where the generator for the dynamics $\mathcal{L}\left(  x_{1},x^{\prime
},t\mid\left\{  \mathbf{R}\left(  t\right)  \right\}  \right)  $ is defined in
terms of higher order time correlation functions. This exact representation is
useful for addressing the question of origin and context of the Kubo-Greenwood model.

An approximation for $\mathcal{L}\left(  x,x^{\prime},t\mid\left\{
\mathbf{R}\left(  t\right)  \right\}  \right)  $ that does not contravene the
conditions of strong Coulomb coupling or other extreme conditions is the
Markov limit
\begin{equation}
\mathcal{L}\left(  x,x^{\prime},t\mid\left\{  \mathbf{R}\left(  t\right)
\right\}  \right)  \rightarrow\mathcal{L}\left(  x,x^{\prime},t=0,\left\{
\mathbf{R}\right\}  \right)  \label{3.4}%
\end{equation}
This means that the generator of the time dependence is the same at all times;
as written, this approximation is exact at $t=0$. At this point we specialize
to the semi-classical limit in which the electrons are treated as classical
particles but the ion - electron Coulomb interaction is regularized at short
distances (see, for example ref \cite{Desjarlais17}). Remarkably, this generator then can be
evaluated without further approximation to give the kinetic equation
\cite{Wrighton08}%
\[
\left(  \partial_{t}+\mathbf{v}\cdot\nabla_{\mathbf{r}}-m^{-1}\nabla
_{\mathbf{r}}\mathcal{V}_{ie}\left(  \mathbf{r},\left\{  \mathbf{R}\right\}
\right)  \cdot\nabla_{\mathbf{v}}\right)  \mathcal{J}(x,\left\{
\mathbf{R}\right\}  ,t)
\]%
\begin{equation}
=-\mathbf{v}\cdot\nabla_{\mathbf{r}}\beta\int dx^{\prime}\mathcal{V}%
_{ee}\left(  \mathbf{r},\mathbf{r}^{\prime},\left\{  \mathbf{R}\right\}
\right)  \phi\left(  v^{\prime}\right)  n\left(  \mathbf{r}^{\prime},\left\{
\mathbf{R}\right\}  \right)  \mathcal{J}(x^{\prime},\left\{  \mathbf{R}%
\right\}  ,t). \label{3.4a}%
\end{equation}
The left side describes the motion of independent particles interacting with
the ions via a "renormalized" potential, found to be%
\begin{equation}
\mathcal{V}_{ie}\left(  \mathbf{r},\left\{  \mathbf{R}\right\}  \right)
=v_{KS}(\mathbf{r},\left\{  \mathbf{R}\right\} \mid n_{e} ), \label{3.6}%
\end{equation}
where $v_{KS}(\mathbf{r},\left\{  \mathbf{R}\right\} \mid n_{e} )$ is the Kohn-Sham
potential defined in terms of the classical electron free energy functional.
This is precisely the dynamics generated by the Kubo-Greenwood model (in its classical limit)!

If these particles were truly independent then the right side of (\ref{3.4a}) would vanish. This is not the case for the kinetic theory here, showing that the Kubo-Greenwood model neglects an
interaction among these particles. The right side of (\ref{3.4a}) describes a
dynamical screening due to the renormalized electron-electron interaction 
\begin{equation}
\mathcal{V}_{ee}\left(  \mathbf{r},\mathbf{r}^{\prime},\left\{  \mathbf{R}%
\right\}  \right)  =-\beta^{-1}c\left(  \mathbf{r},\mathbf{r}^{\prime
},\left\{  \mathbf{R}\right\}  \right), \label{3.7}%
\end{equation}
where $c\left(  \mathbf{r},\mathbf{r}^{\prime},\left\{  \mathbf{R}\right\}
\right)  $ is the electron direct correlation function \cite{Lebowitz69}. In
summary, the Kubo-Greenwood model is seen to be given by the exact short
time dynamics with neglect of the dynamical screening.

\vspace*{-6pt}
\section{Kubo-Greenwood with ion dynamics}\label{sec4}

Return now to the exact Green-Kubo form (\ref{2.3}) and
introduce the Kubo-Greenwood approximation (\ref{2.6}), but without the
additional assumption of static ions
\begin{equation}
H_{e}+U_{ei}\left(  \left\{  \mathbf{R}\left(  t\right)  \right\}  \right)
\rightarrow\sum_{i=1}^{N_{e}}h_{KS}(j,\left\{  \mathbf{R}\left(  t\right)
\right\}  ),\label{4.1}%
\end{equation}%
\begin{equation}
h_{KS}(j,\left\{  \mathbf{R}\left(  t\right)  \right\}  )=\frac{p_{j}^{2}%
}{2m_{e}}+v_{KS}(\mathbf{r}_{j},\left\{  \mathbf{R}\left(  t\right)  \right\}
\mid n_{e}).\label{4.2}%
\end{equation}
Since the Hamiltonian is for independent particles, the correlation function
(\ref{2.3}) can be reduced to a single particle average%
\begin{equation}
\psi_{e}\left(  t,\mid \left\{  \mathbf{R}\left(  t\right)  \right\}  \right)
=\frac{1}{3V}Tr_{1}n\left(  \left\{  1,\mathbf{R}\left(  t\right)  \right\}
\right)  U^{\dagger}\left(  1,t\right)  \mathbf{j}\left(  1\right)  U\left(
1,t\right)  \left(  1-n\left(  1,\left\{  \mathbf{R}\left(  t\right)
\right\}  \right)  \right)  \cdot\mathbf{j}\left(  1\right)  \mathbf{,}%
\label{4.3}%
\end{equation}
where $\mathbf{j}\left(  1\right)  \mathbf{=}e\mathbf{v}_{1}$ and $n\left(
1,\left\{  \mathbf{R}_{k}\left(  t\right)  \right\}  \right)  $ is the Fermi
occupation number operator
\begin{equation}
n\left(  1,\left\{  \mathbf{R}_{k}\left(  t\right)  \right\}  \right)
=\left(  e^{\beta\left(  h_{KS}\left(  1,\left\{  \mathbf{R}_{k}\left(
t\right)  \right\}  \right)  -\mu\right)  }+1\right)  ^{-1}.\label{4.4}%
\end{equation}
The single particle time dependence is obtained from
\begin{equation}
U\left(  1,t\right)  =T\exp\left(  -i\int_{0}^{t}\overline{\epsilon}\left(
\tau\right)  d\tau\right)  ,\label{4.5}%
\end{equation}
where $T$ is a time ordering operator (earlier times to the right). The
quantity $\overline{\epsilon}\left(  \tau\right)  $ has matrix elements in a
representation using the eigenvalues of $h_{KS}\left(  1,\left\{
\mathbf{R}_{k}\left(  t\right)  \right\}  \right)  $%
\begin{equation}
\overline{\epsilon}_{\kappa_{1}\left(  t\right)  ,\kappa_{2}\left(  t\right)
}\left(  \tau\right)  =\sum_{\kappa\left(  \tau\right)  }c\left(  \kappa
_{1}\left(  t\right)  ,\kappa\left(  \tau\right)  \right)  \epsilon
_{\kappa\left(  \tau\right)  }\left(  \left\{  \mathbf{R}\left(  \tau\right)
\right\}  \right)  c\left(  \kappa\left(  \tau\right)  ,\kappa_{2}\left(
t\right)  \right)  .\label{4.5a}%
\end{equation}
The notation $\kappa\left(  t\right)  $ labels the eigenfunction
$h_{KS}(1,\left\{  \mathbf{R}\left(  t\right)  \right\}  )$ with eigenvalue
$\epsilon_{\kappa\left(  t\right)  }\left(  \left\{  \mathbf{R}\left(
t\right)  \right\}  \right)  $. The coefficients $c\left(  \kappa_{1}\left(
t\right)  ,\kappa\left(  \tau\right)  \right)  $ are overlap integrals between
the eigenfunctions of $h_{KS}(1,\left\{  \mathbf{R}\left(  \tau\right)\right\}  )$ and those of
$h_{KS}(1,\left\{  \mathbf{R}\left(  t\right)  \right\}  )$.

Consider first the case of static ions, $\left\{  \mathbf{R}\left(  t\right)
\right\}  \rightarrow\left\{  \mathbf{R}\right\}  $. Then (\ref{4.3}) can be
evaluated using the eigenfunctions and eigenvalues of $h_{KS}(1,\left\{
\mathbf{R}\right\}  )$%
\begin{equation}
\psi_{e}\left(  t,\left\{  \mathbf{R}\right\}  \right)  =\frac{1}{3V}%
{\displaystyle\sum\limits_{\kappa_{1},\kappa_{2}}}
Tr_{1}n_{\kappa_{1}}\left(  \left\{  \mathbf{R}\right\}  \right)  e^{i\left(
\epsilon_{\kappa_{1}}\left(  \left\{  \mathbf{R}\right\}  \right)
-\epsilon_{\kappa_{2}}\left(  \left\{  \mathbf{R}\right\}  \right)  \right)
t}\mathbf{j}_{\kappa_{1}\kappa_{2}}\left(  1-n_{\kappa_{2}}\left(  \left\{
\mathbf{R}\right\}  \right)  \right)  \cdot\mathbf{j}_{\kappa_{2}\kappa_{1}%
}\mathbf{,} \label{4.6}%
\end{equation}%
\begin{equation}
n_{\kappa_{1}}\left(  \left\{  \mathbf{R}\right\}  \right)  =\left(
e^{\beta\left(  \epsilon_{\kappa_{1}}\left(  \left\{  \mathbf{R}\right\}
\right)  -\mu\right)  }+1\right)  ^{-1}. \label{4.7}%
\end{equation}
Here $\epsilon_{\kappa}\left(  \left\{  \mathbf{R}\right\}  \right)  $ is an
eigenvalue of $h_{KS}(1,\left\{  \mathbf{R}\right\}  )$. These are the Kubo-Greenwood results.

More generally, with $\left\{  \mathbf{R}\left(  t\right)  \right\}  $,
(\ref{4.3}) can be evaluated using the eigenfunctions of $h_{KS}(1,\left\{
\mathbf{R}\left(  t\right)  \right\}  )$ as a basis set
\begin{align}
\psi_{e}\left(  t,\left\{  \mathbf{R}\left(  t\right)  \right\}  \right)   &
=\frac{1}{3V}%
{\displaystyle\sum\limits_{\kappa_{1},\kappa_{2}}}
Tr_{1}n_{\kappa_{1}\left(  t\right)  }\left(  \left\{  \mathbf{R}\left(
t\right)  \right\}  \right)  U_{\kappa_{1}\left(  t\right)  \kappa_{4}\left(
t\right)  }^{\dagger}\left(  t\right)  \mathbf{j}_{\kappa_{4}\kappa_{3}%
}U_{\kappa_{3}\left(  t\right)  \kappa_{2}\left(  t\right)  }\left(  t\right)
\nonumber\\
&  \times\left(  1-n_{\kappa_{2}\left(  t\right)  }\left(  \left\{
\mathbf{R}\left(  t\right)  \right\}  \right)  \right)  \cdot\mathbf{j}%
_{\kappa_{2}\left(  t\right)  \kappa_{1}\left(  t\right)  }\mathbf{.}%
\label{4.8}%
\end{align}
The evaluation of $U_{\kappa_{3}\left(  t\right)  \kappa_{2}\left(  t\right)
}\left(  t\right)  $ now is somewhat more complex as it entails the overlap
integrals $c\left(  \kappa_{1}\left(  t\right)  ,\kappa\left(  \tau\right)
\right)  $ for all $\tau<t.$ In practice, the time evolution of the ions is
provided by AIMD on a discrete set of times - the ion time step. Then
$U_{\kappa_{3}\left(  t\right)  \kappa_{2}\left(  t\right)  }\left(  t\right)
$ can be written as an ordered product for each time interval. If the time
correlation function $\psi_{e}\left(  t,\left\{  \mathbf{R}\left(  t\right)
\right\}  \right)  $ decays sufficiently rapidly, only a few members of this
product might be required. For any finite time dependence for the ions, the
time integral of (\ref{2.1}) will entangle the electron and ion dynamics so
that the usual Kubo-Greenwood form will no longer hold.

\section{Discussion}\label{sec6}

The analysis summarized in Section (\ref{sec3})  provides partial support for the Kubo-Greenwood phenomenology (origin of the assumed independent single particle dynamics based on the equilibrium Kohn-Sham Hamiltonian) and its context (nature of the corrections due to interactions among these "quasi-particles"). The latter corrections can be incorporated through exact solution to the kinetic equation  (\ref{3.4a}), giving rise to a random phase approximation form with a Kubo-Greenwood polarization function, and electron-electron interactions replaced by(\ref{3.7}). Still missing in this kinetic equation are the effects of electron-electron collisions (ion-electron collisions are in fact accurately described by (\ref{3.4a})). The effects of these collisions can be added via a single relaxation time model collision operator. Their importance has been the subject of recent investigations \cite{Desjarlais17,Reinholz15}.

The effects of ion motion during the single particle Kohn-Sham electron dynamics are described in Section  
(\ref{sec4}). The inclusion of these effects, while still within the context of Kohn - Sham calculations, compromises some of the simplicity of the Kubo-Greenwood model. It is expected that the time scales of the electron and ion dynamics differ by roughly the square root of their mass ratio. It remains to determine how rapidly the electron averaged correlation function $\psi_{e}\left(  t\mid\left\{  \mathbf{R}\left(  t\right)
\right\}  \right)  $ decays in order to assess the importance of ion dynamics.

\section{Acknowledgments}

This research has been supported by US DOE Grant DE-SC0002139.

\end{document}